\documentclass[letter]{aa} % for the letters 
\usepackage{xcolor}

\usepackage{subcaption}   
\usepackage{float}
\usepackage{graphicx}
\usepackage{txfonts}
\usepackage{mathptmx}
\usepackage{placeins}
\usepackage[colorlinks=true,linkcolor=blue,allcolors=blue]{hyperref}
\hypersetup{colorlinks,
citecolor=blue
}

\newcommand{\powerlaw}{{\fontfamily{qcr}\selectfont PL}}

\newcommand{\gabs}{{\fontfamily{qcr}\selectfont GABS}}

\newlength{\imagew}
\newlength{\imageh}
\newlength{\legendw}
\newlength{\legendh}
\newlength{\legendx}
\newlength{\legendy}
\newcommand{\graphicswithlegend}[6]{
  \setlength{\imagew}{#1}
  \settoheight{\imageh}{\includegraphics[width=\imagew]{#2}}

  \setlength{\legendw}{#3\imagew}
  \settoheight{\legendh}{\includegraphics[width=\legendw]{#4}}

  \setlength{\legendx}{\imagew}
  \addtolength{\legendx}{-\legendw}
  \addtolength{\legendx}{-#5\imagew}

  \setlength{\legendy}{\imageh}
  \addtolength{\legendy}{-\legendh}
  \addtolength{\legendy}{-#6\imageh}

  \includegraphics[width=\imagew]{#2}%
  \llap{
    \hspace{-\the\legendx}
    \raisebox{\legendy}{\includegraphics[width=\legendw]{#4}}
    \hspace{\the\legendx}
  }
}

\begin{document}

\title{Feeding and Feedback in Dwarf Galaxies (FeeD) - I. Evidence of nuclear ultra-fast and galaxy-scale outflows in the dwarf galaxy Arp 151}

\titlerunning{Outflows in the dwarf galaxy Arp 151}

\author{Santanu Mondal
          \inst{1,2}
          \and
          Ankit Patel\inst{2}
          \and
          Mar Mezcua\inst{3,4}
          \and
          Ravi Joshi\inst{2}
          \and
          Yerong Xu\inst{3,4}
          \and
          K. Aditya\inst{5}
          \and
          Smitha Subramanian\inst{2}
          \and
          Victor Rodriguez Morales\inst{3}
          }

\authorrunning{Mondal et al.}

   \institute{Department of Physics and Electronics, Christ University, Bangalore 560029, India, 
   \email{santanuicsp@gmail.com}
   \and
   Indian Institute of Astrophysics, II Block, Koramangala, Bangalore 560034, India,     
   \and  
    Institute of Space Sciences (ICE, CSIC), Campus UAB, Carrer de Can Magrans, s/n, 08193 Barcelona, Spain
    \and
    Institut d’Estudis Espacials de Catalunya (IEEC), Edifici RDIT, Campus UPC, 08860 Castelldefels, Barcelona, Spain
    \and 
    Raman Research Institute, C. V. Raman Avenue, 5th Cross Road, Sadashivanagar, Bengaluru, 560080, India}

\abstract 
{Feeding and feedback regulated by supermassive black holes (SMBHs) play a central role in galaxy growth and evolution, yet these processes remain poorly understood in low-mass galaxies. In particular, the presence, properties, and role of ultra-fast nuclear outflows (UFOs) in low-mass galaxy systems are largely unexplored. We analyze available NuSTAR X-ray observations of Arp 151 and find a possible evidence ($\sim 2\sigma$ confidence) for a fast outflow with a velocity of $\sim0.18c$ from the central BH. Furthermore, we have also detected an optical galaxy-scale outflow in MaNGA Integral Field Unit data. The estimated nuclear and galaxy-scale mass outflow rates are $\sim0.015$ $M_\odot$/yr from {\it NuSTAR} and $\sim0.43$ $M_\odot$/yr from MaNGA, respectively. Our estimates suggest that such outflows may significantly regulate the feedback process in the galaxy. Comparing the kinetics of the UFO and the galaxy-scale outflow indicates that they are in the momentum-conserving phase. This tentative detection implies that dwarf galaxies are also able to generate UFOs, which so far have been detected in massive galaxies. Thus, the AGN feedback may also be important for the evolution of the dwarf galaxies. }


\maketitle

\section{Introduction}

Active galactic nuclei (AGN) are powered by accretion of matter onto supermassive black holes (SMBHs; $\sim10^{6-10} M_\odot$) at the centre of massive galaxies. During this process, a large amount of energy is released in the form of radiation, relativistic jets, and outflows that can interact with the interstellar medium of the host galaxy. This is commonly known as AGN feedback, which can heat or expel gas from the galaxy, thereby regulating star formation and influencing the growth of both the SMBH and its host galaxy \citep{SilkRees1998A&A...331L...1S,Fabian2012ARA&A..50..455F}.

The effect of feeding and feedback mechanisms is ill-understood for massive BHs ($10^5 - 10^6$ M$_\odot$) in dwarf galaxies (with stellar mass $<10^{9.5} M_\odot$).  
Compared to massive galaxies, AGN in dwarf systems tend to be fainter and less powerful \citep[see,][]{MezcuaEtal2024ApJ...966L..30M}.  
The AGN feedback may play a crucial role in regulating star formation in dwarf galaxies and is an important testbed of galaxy formation and cosmology \citep[][and references therein]{KoudmaniEtal2021MNRAS.503.3568K}. Furthermore, if the BHs in these systems grow efficiently, they may have more effect on the host galaxy \citep{BaraiDalPino2019MNRAS.487.5549B,KoudmaniEtal2021MNRAS.503.3568K}. Recent studies found direct evidence of spatially extended high-velocity ionized gas outflows in dwarf galaxies consistent with AGN-driven outflows \citep[][]{Manzano-kingEtal2019ApJ...884...54M,MoralesEtal2025A&A...697A.235R,SalehiradEtal2025ApJ...979...26S}. 

There is a growing number of evidence for the presence of (sub)relativistic ($\gtrsim 0.1c$) outflows in the X-ray spectra of AGN in massive galaxies, imprinted by blue-shifted Fe K-shell absorption lines from Fe XXV and/or Fe XXVI at rest-frame energy $>7$ keV \citep[][and references therein]{PapadakisEtal2007A&A...461..931P,SimEtal2010MNRAS.408.1396S,TombesiEtal2010A&A...521A..57T,MatzeuEtal2023A&A...670A.182M}. These lines have been observed in a highly-ionized medium with column density $\geq 10^{22}$ cm$^{-2}$ and blue-shifted velocity $\gtrsim 0.2c$ \citep[][]{PoundsEtal2003MNRAS.345..705P,Cappi2006AN....327.1012C,BraitoEtal2007ApJ...670..978B}, and are known as ultra-fast outflows (UFOs). The kinetic power of UFOs can significantly influence the host galaxy environment and thus their evolution \citep[][and references therein]{TombesiEtal2013MNRAS.430.1102T}, consistent with the SMBH-galaxy feedback driven by outflows \citep[][]{KingPounds2003MNRAS.345..657K,GaspariEtal2011MNRAS.411..349G}. 

However, no such evidence of UFOs from the inner central engine has been reported in dwarf galaxies and the regulator of the feeding and feedback process. These massive BHs might be the counterparts of the seed BHs formed in the early universe \citep[][]{MezcuaEtal2023ApJ...943L...5M}. The SMBH mass and accretion rate scales the energetics and mass outflow rates of these UFOs  \citep{KingPounds2003MNRAS.345..657K,TombesiEtal2013MNRAS.430.1102T,MatzeuEtal2023A&A...670A.182M}.

Arp 151 (Mrk 40) is a nearby (z=0.021) dwarf galaxy \citep[with a stellar mass of $10^{9.3} M_\odot$;][]{MezcuaEtal2020ApJ...898L..30M}. It hosts an active SMBH along with two companion galaxies at the same redshift, which indicates that AGN is likely triggered by the merger event. Optical spectroscopic analyses estimated the BH masses in Arp 151 to be $\sim4-7\times 10^6 M_\odot$  \citep[][and references therein]{BentzEtal2008ApJ...689L..21B,ValentiEtal2015ApJ...813L..36V}.

The crucial role of nuclear outflows in the feeding and feedback process and regulating the galaxy growth motivates us in this letter to investigate the multiscale outflows in Arp 151. 
As an initiative of understanding feeding and feedback in dwarf galaxies (FeeD), here, we present the first detailed X-ray and integral field unit (IFU) spectroscopic analysis of outflows in the dwarf galaxy Arp 151 using available archival {\it NuSTAR} and MaNGA data, respectively \citep[see also,][]{VasudevanEtal2013ApJ...763..111V}. The paper is organized as follows: in Sect. 2, we present the X-ray and optical analyses, UFO and galaxy-scale outflow properties. We discuss the results, the impact of UFOs in the galaxy feedback, and draw our conclusion in Sect. 3. 

\begin{figure*}[h!]
\centering{
\includegraphics[height=6.0truecm,width=4.0truecm,angle=270,trim={0.0cm, 0.0cm, 0.0cm, 0.0cm}, clip]{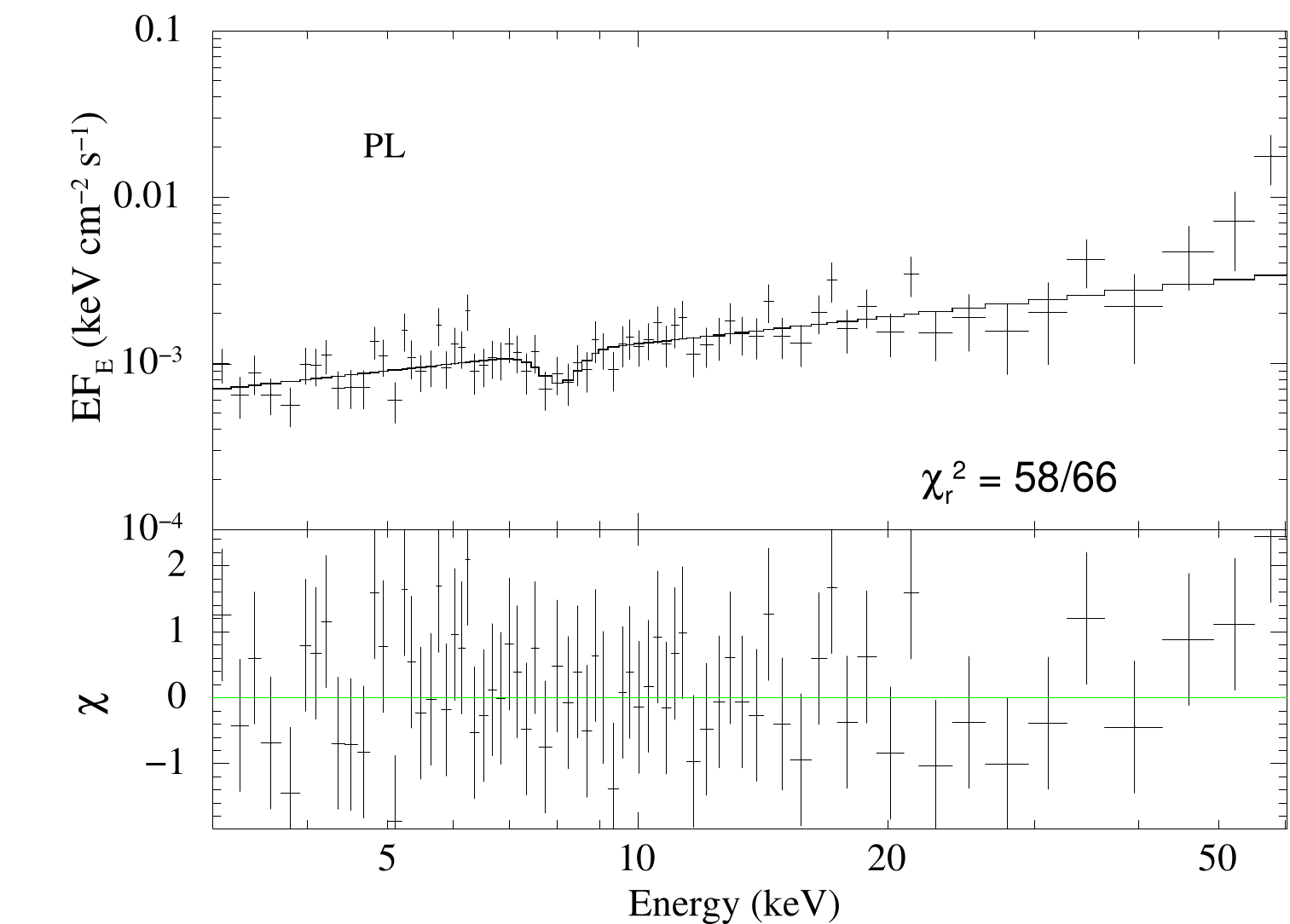}
\includegraphics[height=6.0truecm,width=4.0truecm,angle=270,trim={0.0cm, 0.0cm, 0.0cm, 0.0cm}, clip]{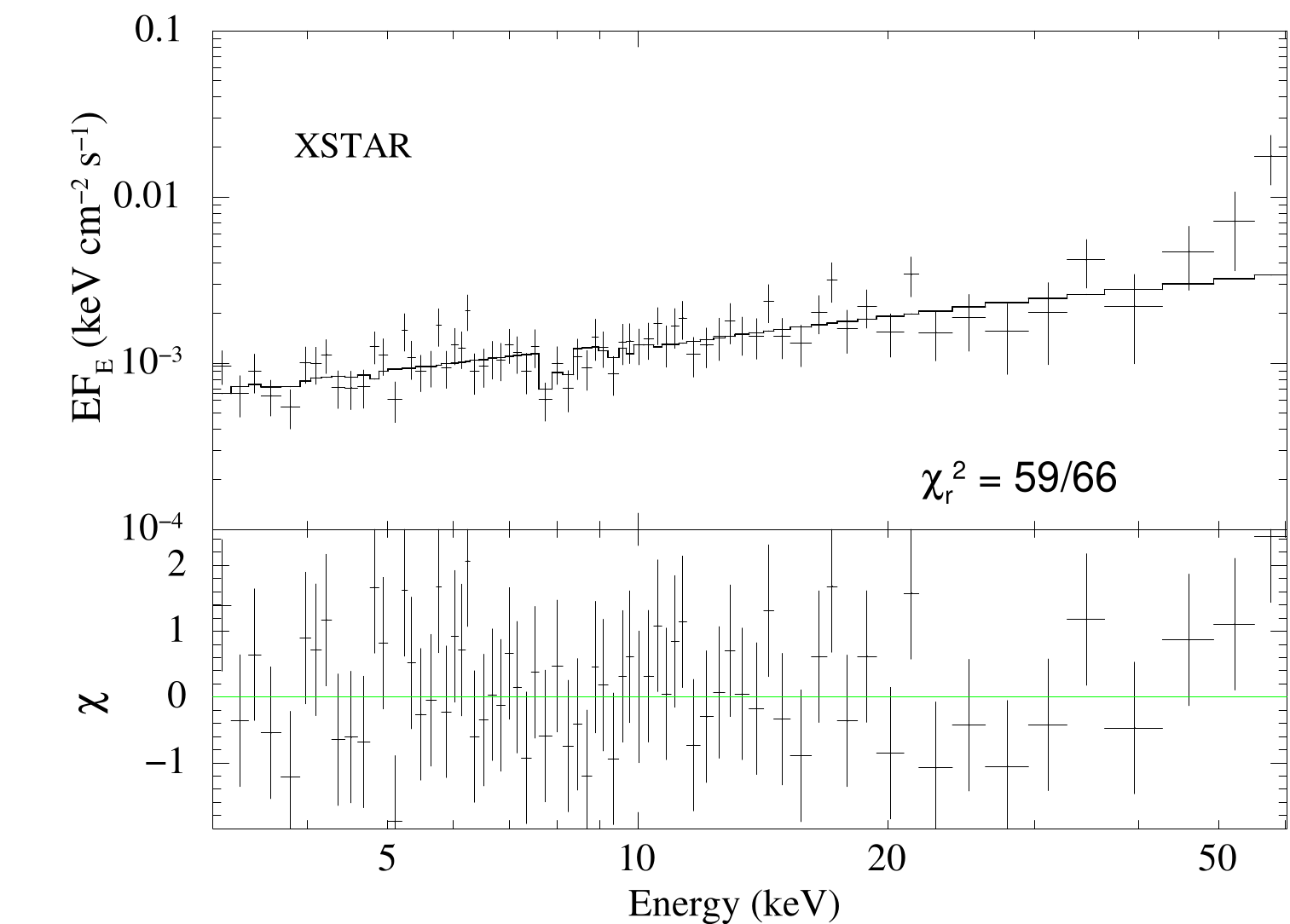}
\includegraphics[height=6.0truecm,width=4.truecm,angle=270,trim={1.0cm, 0.0cm, 0.0cm, 0.0cm}, clip]{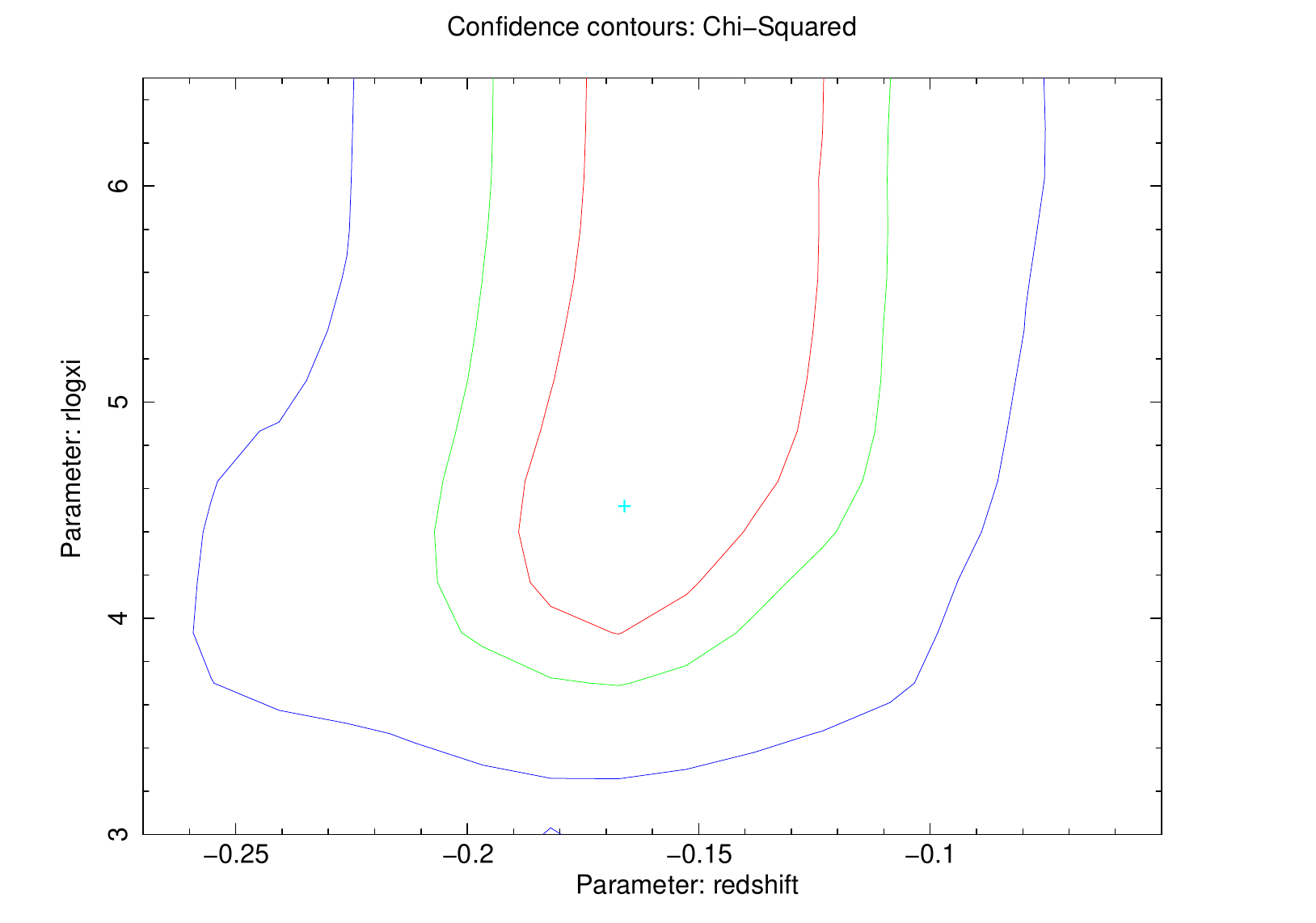}
}
\caption{Left: Best-fitted FPMA spectrum of Arp 151 using the \powerlaw\,model. Middle: The same spectrum fitted using the {\sc XSTAR} model. The solid lines are for the model, and the points with error bars represent the data. The bottom panels show the residuals of the fit. Right: The confidence contour between the wind redshift and ionization parameters. The red, green, and blue lines show the confidence contour in 1, 2, and 3$\sigma$ levels.
} 
\label{fig:spectra}
\end{figure*}

\section{Sub-parsec to galaxy-scale outflows}
\subsection{Ultra-fast outflow}

The X-ray spectroscopic analysis is performed using only {\it NuSTAR} observations from October 8, 2019, with an exposure time of 24 ks (observation ID 60160430002). No data is available from other X-ray missions. We used standard {\tt nupipeline} and {\tt nuproducts} tasks to extract the spectrum. A source region of $40^{\prime\prime}$ and a background region of $80^{\prime\prime}$ are considered in DS9 to extract the spectrum. The source FPMA and FPMB spectra are grouped with a minimum of 15 counts in each bin using the {\tt grppha} task. The spectral modeling for the energy band 3-60 keV is performed in {\tt XSPEC} (v12.12.1) \citep[][]{Arnaud1996} using both phenomenological and physical models. Galactic absorption of $N_H=1.1\times10^{20}$ cm$^{-2}$ \citep[][]{VasudevanEtal2013ApJ...763..111V} is used throughout the analysis.

We first fitted the data using a simple power-law (\powerlaw)\,model along with the line energy shift {\tt ZASHIFT} and Gaussian absorption \gabs\, model for the absorption line. The best-fit with C-stat 67 is achieved for the powerlaw model (\powerlaw) photon index, $\Gamma=1.46\pm0.06$. The inclusion of \gabs\,improved the fit by reducing $\Delta$C-stat by 9 for a loss of three degrees of freedom, which is equivalent to a $\sim2\sigma$ detection using F-test probability. This significance does not account for the look-elsewhere effect \citep{ProtassovEtAl2002ApJ...571..545P,GrossVitells2010EPJC...70..525G}. Interestingly, we observed a tentative blueshifted absorption line at energy $E_{gabs}=8.08\pm0.24$ keV (observer frame) with line width $\sigma_{gabs}=0.41\pm0.27$ keV and line strength $0.45\pm0.22$. The rest-frame energy of the line is $8.25\pm0.24$ keV for $z=0.021$ in the {\tt ZASHIFT} model. Such absorption lines may originate due to high-velocity outflowing winds, also known as UFOs detected in preferentially massive AGN \citep[see,][for detailed discussions]{PapadakisEtal2007A&A...461..931P,TombesiEtal2010A&A...521A..57T}. The left panel of \autoref{fig:spectra} shows the best-fitted model spectra. This phenomenological modeling can give us the preliminary information about the spectral shape and absorption line properties. Such a line feature can be identified with highly ionized iron absorption lines, associated with Fe XXV or Fe XXVI K-shell transitions. The most plausible assumption is that the observed $8.25$ keV line can be the blue-shifted line corresponding to Fe XXV or Fe XXVI K$\alpha$ at a rest energy of 6.97 keV. If there is an outflowing wind aligned with our line-of-sight (LOS), a line can be blue-shifted. Under that assumption, we calculated the wind velocity using the special relativistic Doppler shift formula $v_w=c [(E/E_{gabs})^2-1]/[(E/E_{gabs})^2+1]$, which yields $v_w\sim0.17c$ for $E=6.97$ keV, and $E_{gabs}=8.25$ keV, respectively. We performed the same analysis using FPMB data; however, we could not detect a clear line feature. 

To further understand the properties of the outflowing material, we have refitted the data using the more realistic photoionization model XSTAR  \citep{KallmanBautista2001ApJS..133..221K}, which can model the absorption line and its properties. For our fitting purposes, we have frozen the turbulence velocity ($v_{\rm turb}$) to a typical 2000 km s$^{-1}$ \citep[][]{Xrism2025} and generated the table model for $\Gamma=1.5$. The best-fit is achieved with C-stat 68, comparable with the phenomenological \gabs\,model. The best-fitted parameters are $N_H=2.61^{+6.76}_{-1.23}\times10^{23}$ cm$^{-2}$, ionization parameter $\log\xi=4.5^{+2.0}_{-0.4}$ (the upper hard limit is set to 6.5), and redshift of the outflow ($z_{abs}$) $-0.17^{+0.02}_{-0.01}$. The middle panel of \autoref{fig:spectra} shows the best-fitted spectrum, and the right panel shows the confidence contour between $\log \xi$ and $z_{\rm abs}$ parameters. The red, green, and blue contours show the 1, 2, and 3$\sigma$ confidence levels, respectively. The contour indicates that the upper limit of $\log \xi$ is not currently constrained, which may require a higher exposure or signal-to-noise data. The best-fitted $z_{\rm abs}$ infers the velocity of the outflow is $0.18c$, remarkably consistent with the estimation from the Doppler shift. The high $N_H$ and $\xi$ appear to favor the Fe XXVI K$\alpha$ identification originated from a high velocity UFO from the inner region of the disk. The distance of the outflow gas can be computed from the best-fit parameters as follows. The minimum distance of the outflowing gas is the radius at which the outflow velocity is equivalent to the escape velocity, $r_{\rm min}=2 G M_{\rm BH}/v_{\rm UFO}^2=6.4\times10^{13}$ cm $=30$ r$_S$, for a BH mass of $7.0\times10^6 M_\odot$. Given the above model parameters, the lower limit of the mass outflow rate can be estimated as $\dot M_{\rm UFO}=\Omega N_H m_p v_{\rm UFO} r_{\rm min}\sim0.015 ~M_\odot yr^{-1}$, where a solid angle $\Omega=2\pi$ is considered \citep[see,][]{NardiniEtal2015Sci...347..860N}. We have further estimated the kinetic power $\dot E_{\rm kUFO} = 1/2 \dot M_{\rm UFO} v_{\rm UFO}^2 \sim 1.4\times 10^{43}$ erg s$^{-1}$ and momentum flux rate $\dot P_{\rm UFO}=5.1\times10^{33}$ gm cm s$^{-2}$.

\subsection{Galaxy-scale outflow}
In this section, we investigate the galactic-scale gas outflows (OF) by analyzing the optical integral field spectroscopy data cube from the MaNGA\footnote{https://magrathea.sdss.org/marvin/galaxy/9000-1901/} survey. 

\subsubsection{Outflow Velocity}
\begin{figure}
     \centering
     \includegraphics[height=6.5truecm,width=8.0truecm]{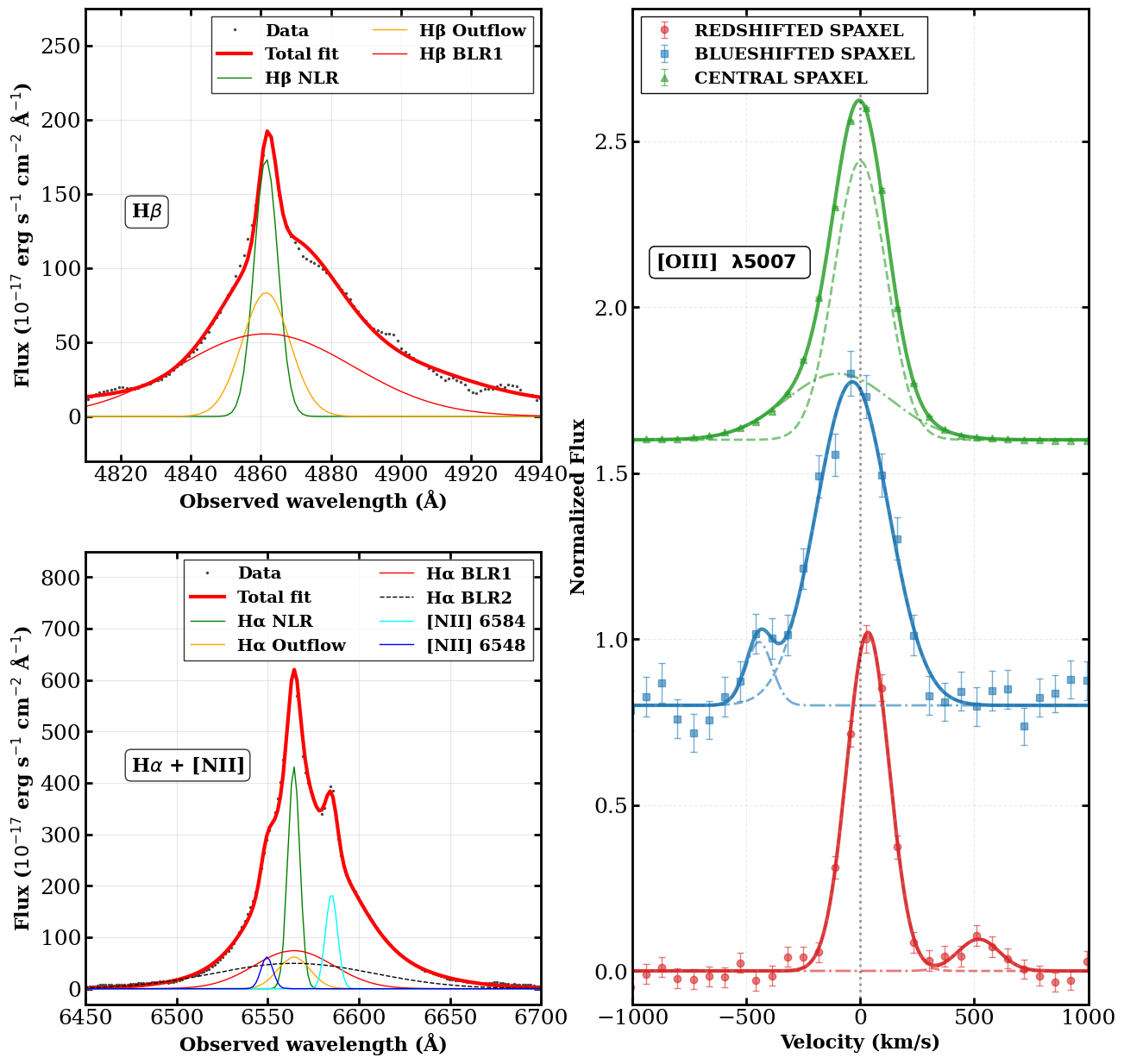}
     \caption{Multiple Gaussian component spectral line modelling of H$\beta$ (top-left) and H$\alpha$ (bottom-left) emission line profiles.  Right panel: The [O III]$\lambda$5007 nebular emission line modelled with two Gaussian components, for the central spaxel with a weak flow signature and additional two spaxels showing strong blue- and red-shifted emission line profiles. The solid line represents model fit, while the dotted lines represent the individual underlying components.}
     \label{fig:balmer_fit}
     
 \end{figure}

\begin{figure}[!h]
\centering{
\includegraphics[height=6.5truecm,width=8.0truecm,trim={0.0cm, 0.0cm, 0.0cm, 0.0cm}, clip]{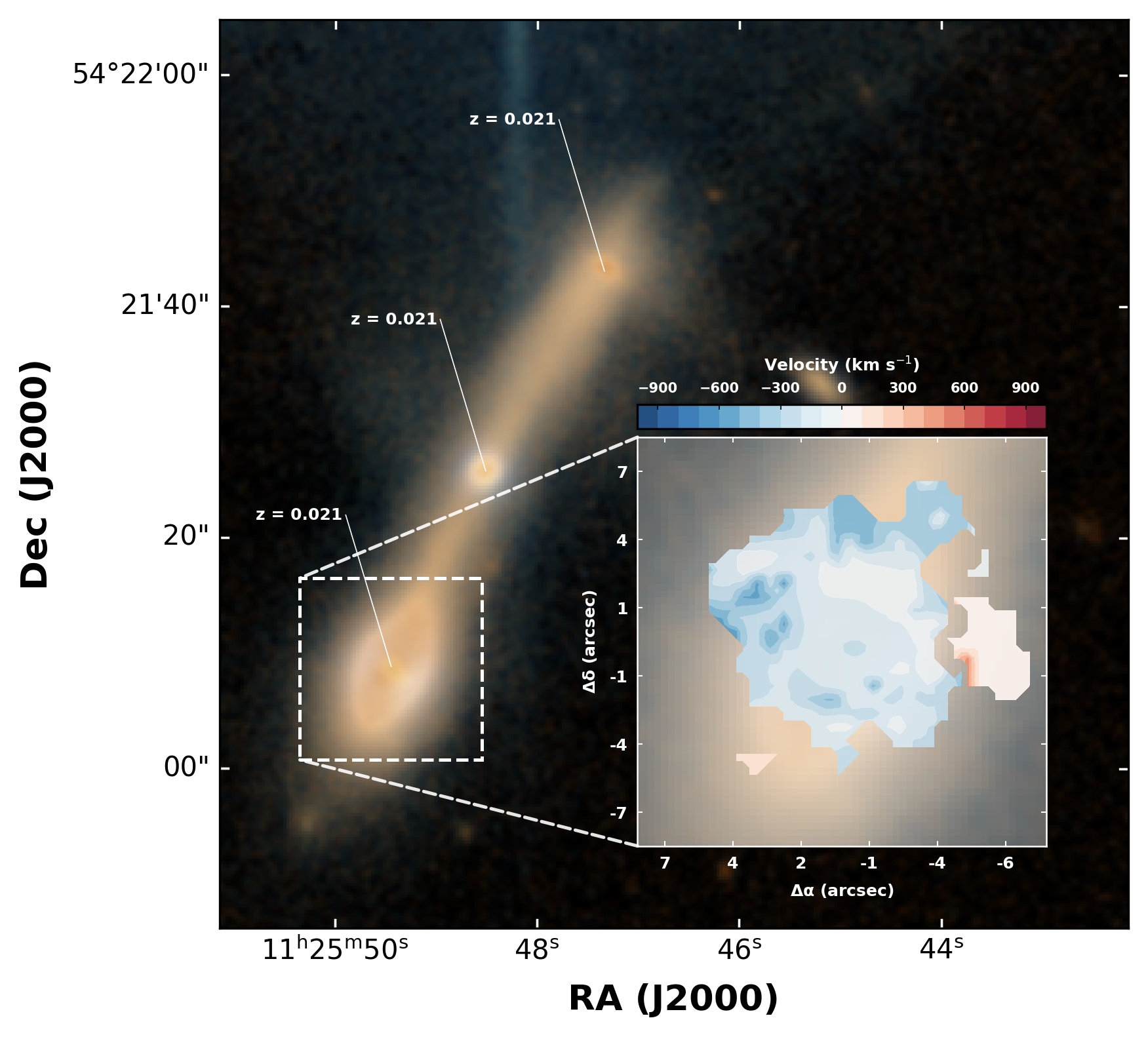}}
\caption{The DECaLS color composite ($g, r,z$-band) image of the dwarf AGN host galaxy Arp 151. Two companion galaxies having a redshift of $z=0.021$ consistent with  Arp 151 indicate the presence of a merger system. The inset shows a bi-conical outflow from the dwarf galaxy.
} 
\label{fig:MangaMap}
\end{figure}

The extent and kinematics of ionized gas outflows are traced using the [O\,\textsc{iii}] $\lambda5007$ emission line, which is an excellent tracer of diffuse ionized gas. For this, we first mask the prominent emission features and model the stellar continuum in each spaxel by using the publicly available non-linear optimization spectral fitting routine \texttt{PyQSOFit} \citep{2018ascl.soft09008G}. We then fit the [O\,\textsc{iii}]$\lambda5007$ emission line using a narrow Gaussian component with a typical width of $\sigma \lesssim 500~\rm km\ s^{-1}$ \citep{2025ApJ...990..231P}, representing the emission from the narrow line region. In addition, a broader component for possible outflow signature is introduced for the spaxels where the F-test yields a significantly better fit compared to the single-Gaussian model, with a confidence level of 95\%. We further verified this by ensuring a lower Bayesian Information Criterion value \citep{1978AnSta...6..461S,2007MNRAS.377L..74L} for the double Gaussian profile as compared to the single Gaussian model. Using the outflow component, we estimated the typical outflow velocity width ($\rm W_{80}$), comprising the 80\% of the flux of the outflow component. In addition, the velocity is estimated as $v_{OF} = - v_{0} + \frac{W_{80}}{2}$, where $v_{0}$ is the velocity offset between the center of the narrow and the outflow component. The [N II] and [S II] doublets were modelled with narrow Gaussian components whose kinematics were tied to the narrow component of [O III] line.  The complex profile of the H$\beta$ and H$\alpha$ line is fitted with multiple (1-4) Gaussians with an initial guess of two narrow, one broad and one very broad component. For a physically consistent fit, the width of the narrow components was tied to the [O III] systemic and outflow features. To reduce the arbitrariness of the component fits and make the decomposition more physical, we have constrained the redshift and width of the two narrow components to be the same as those of the [O~III]$\lambda 5007$ lines. The best-fitted model and components for H$\beta$ and H$\alpha$ emission lines are shown in the left panel of \autoref{fig:balmer_fit}. The FWHM of the H$\beta$ broad emission line component is  $3631 \pm 717$ km s$^{-1}$ measured from the central region (25 spaxels) where the AGN emission dominates, which gives M$_{BH}$ of $6.0\pm0.2 \times 10^6$ M$_\odot$ \citep[see][for the scaling relation]{greene2005estimating}, consistent with other works \citep{BentzEtal2008ApJ...689L..21B,ValentiEtal2015ApJ...813L..36V} that followed a similar method. The inset in \autoref{fig:MangaMap}, shows the velocity map, with an apparent biconical outflowing component seen as blueshifted and redshifted profiles (see also, Figure 2, right panel).

\subsubsection{Mass outflow rate}
The outflowing ionized gas mass ($M_{\rm OF}$) is estimated using the following relation \citep{2024A&A...685A..99C},

\begin{equation}
M_{\rm OF} =
0.8 \times 10^{8}
\left( \frac{L_{\rm [O\,III], OF}}{10^{44} \rm{erg\ s^{-1}}} \right)
\left( \frac{500\ \rm{cm^{-3}}}{n_e} \right)
\left( \frac{Z_\odot}{Z} \right)
M_\odot .
\end{equation}
Where $L_{\rm [O\,III], OF}$ is the luminosity of the outflowing component of [O III] $\lambda5007$ line, $n_e$ the electron density, and $Z$ is the gas phase metallicity. The electron number density is calculated as $n_e = (Cr - AB)/(A - r)$ \citep[][]{2016ApJ...816...23S}. Here $r$ is the flux ratio of the [S\,\textsc{ii}] emission lines, defined as
$r = \rm{[S\,II]}\,\lambda6716 / \rm{[S\,II]}\,\lambda6731$ and $A$, $B$ and $C$ are dimensionless constants equal to 0.4314, 2107, and 627.1, respectively. The gas phase metallicity is calculated as $Z/Z_\odot = 4.01^{\rm{N2} - 0.07}$, where
$\rm{N2} = \log\left([\rm{N\,II}]\,\lambda6584/\rm{H}\alpha\right)$ \citep[see also,][]{10.1093/mnras/staa193}. The estimated mean electron density, $r$ and metallicity values are 356 cm$^{-3}$, 1.2162 and 0.64 Z$_\odot$, respectively.

Using the above 2D outflow map, we obtained the mass outflow rate as $\dot{M}_{\rm{OF}} = v_{\rm{OF}}\, M_{\rm{OF}}/R_d$,
where $R_d$ is the distance of each spaxel hosting outflow with respect to the galaxy center.  We detect the extended outflow imprints with a physical size of $\sim 2$kpc (4 arcsec) from the galactic center. The integrated $\dot M_{\rm OF}$ and maximum $v_{\rm OF}$ is found to be an average 0.43 M$_\odot$ yr$^{-1}$ and $\approx \pm1000$ km s$^{-1}$ respectively. In massive galaxies $\dot M_{\rm OF} \sim$ a few $\times 10 M_\odot$ yr$^{-1}$ and $v_{\rm OF} \sim$ a few$\times10^3$ cm s$^{-1}$  are significantly higher in magnitude, which suggests that outflows in dwarf galaxies are possibly the scale-down version of those in massive galaxies \citep{OsterbrockFerland2006agna.book.....O,AveryEtal2021MNRAS.503.5134A,MoralesEtal2025A&A...697A.235R}. We have estimated the kinetic energy of the mass outflow in galaxy-scale as $\dot E_{\rm kOF}=1/2 \dot M_{\rm OF}v_{\rm OF}^2=1.4\times10^{41}$ erg s$^{-1}$ and the momentum flux $\dot P_{OF}=\dot M_{\rm OF}v_{\rm OF}=2.7\times10^{33}$ g cm s$^{-2}$. 

To quantify the impact of the UFO and galaxy-scale outflow on the host galaxy, we have estimated the AGN bolometric luminosity ($L_{\rm bol}$) using the relation $L_{\rm bol}=1000\times L_{\rm[OIII]}$ \citep{MoranEtal2014AJ....148..136M}. The $L_{\rm [OIII]}$ measured over the central 25 spaxels, corresponding to a typical seeing of $\sim$2.5 arcsec, the primary AGN emission at the core, results in $L_{\rm [OIII]}$ = $4.1\times10^{40}$ erg s$^{-1}$ and $L_{\rm bol}=4.1\times10^{43}$ erg s$^{-1}$. The radiation momentum flux is found to be $\dot P_{\rm rad}=L_{\rm Edd}/c=2.4\times 10^{34}$ g cm s$^{-2}$ for the Eddington luminosity of $L_{\rm Edd}$ = $7.2\times 10^{44}$ erg s$^{-1}$ for a BH mass $\sim6\times 10^6 M_\odot$ taken from this work.

\section{Discussion and conclusions}
The UFOs launched from the sub-parsec region of the central engine propagate through the interstellar medium and leave their imprint on the galaxy-scale gas dynamics. During this process, they deposit both energy and momentum to the surrounding medium and can potentially affect the star formation of the host galaxy \citep{KingPounds2003MNRAS.345..657K,DiMatteoEtal2005Natur.433..604D,GaspariEtal2011MNRAS.411..349G,Quataert2012MNRAS.425..605F}.In this work, we have analyzed the X-ray data obtained from {\it NuSTAR} and the optical data from MaNGA of the nearby dwarf galaxy Arp 151 to understand the sub-pc to galaxy-scale outflows and their implications for the host galaxy. 
We have tentatively evidenced a UFO signature in the X-ray spectrum based on the detection of an absorption line at $\sim 8.08$ keV (observer frame) with $\sim2\sigma$ confidence. If the absorption line corresponds to Fe XXVI K$\alpha$, it suggests the presence of a highly ionized UFO with $N_H\sim2.6\times10^{23}$ cm$^{-2}$ moving away from the central source with a velocity $\sim0.18c$. The combined X-ray and optical analyses raise the question of whether such UFOs in dwarf galaxies are able to produce sufficient feedback. An important parameter that quantifies the feedback contribution of outflows in AGN is the ratio of the kinetic power of the outflow and the bolometric luminosity, $\epsilon_{UFO}$ \citep{DiMatteoEtal2005Natur.433..604D,GaspariEtal2011MNRAS.411..349G}. For Arp 151, we have estimated $\epsilon_{\rm UFO}=\dot E_{\rm kUFO}/L_{\rm Edd}\sim0.02$, i.e., the $\dot E_{\rm kUFO}$ is 2 percent of $L_{\rm Edd}$. This $\epsilon_{\rm UFO}$ may suggest significant feedback by the UFO to the host galaxy. The momentum load of the nuclear outflow is $\dot P_{\rm UFO}/\dot P_{\rm rad}=0.2$, less compared to the radiatively driven winds or outflows from the accretion disk \citep[$\sim 1$ in][]{KingPounds2003MNRAS.345..657K}. 

Moreover, we compare the energy budgets of the galaxy-scale outflow and the UFO. The corresponding $\epsilon_{\rm OF}=\dot E_{\rm kOF}/L_{\rm Edd}\sim 0.002$ and $\dot P_{\rm OF}/\dot P_{\rm rad}=0.1$ are less compared to the requirements (0.5-5\%) typically assumed by the most popular feedback models \citep{DiMatteoEtal2005Natur.433..604D,GaspariEtal2011MNRAS.411..349G}. Such a scenario may appear if the cold gas or ionized outflows capture only a tiny fraction of the UFO energy, and most of the UFO energy stays in a hot shocked phase while interacting with the ambient medium \citep[][for a review]{Quataert2012MNRAS.425..605F,FioreEtal2017A&A...601A.143F}. This can be further verified from the estimation of $\dot E_{\rm kOF}/E_{\rm kUFO}$ and $\dot P_{\rm OF}/\dot P_{\rm UFO}$. These ratios come out to be $\approx 0.01$ and $\approx 0.6$, which says that neither the energy nor the momentum of the UFO has totally transferred to the galactic scale warm gas; therefore, the galactic scale outflow may be in a momentum-conserving phase. 

Our estimated feedback parameter $\epsilon_{\rm UFO}\sim0.02$ suggests a significant impact of the UFOs on the host galaxy's feedback mechanism. Such activities have been observed in massive galaxies when they host an AGN. However, no such effects have been reported in the case of dwarf galaxies to date. To further understand the impact of the UFO on the galaxy-scale outflows, we have estimated their ratios of energy and momentum rates, and found that both the UFO and the galaxy-scale outflows might be in the momentum-conserving phase.

\begin{acknowledgements}
MM acknowledges support from the Spanish Ministry of Science and Innovation through the project PID2024-159201NB-C22 and also partly supported by the Spanish program Unidad de Excelencia Mar\'ia de Maeztu CEX2020-001058-M, financed by MCIN/AEI/10.13039/501100011033, and by the MaX-CSIC Excellence Award MaX4-SOMMA-ICE. VRM acknowledges support from the Spanish Ministry of Science, Innovation and Universities through the project PRE2022-104649. This research has utilized the {\it NuSTAR} Data Analysis Software ({\sc nustardas}), jointly developed by the ASI Science Data Center (ASDC), Italy, and the California Institute of Technology (Caltech), USA.
\end{acknowledgements}

\bibliographystyle{aa}
\bibliography{example.bib}

%\end{appendix}

\end{document}